\documentstyle[12pt]{article}
\newcommand{\Dslash}{\not{\hbox{\kern-4pt $D$}}}
\newcommand{\Tr}{\mathop{\rm Tr}}
\newcommand{\tr}{\mathop{\rm tr}}
\begin{document}

\begin{flushright}{UT-03-13\\IU-MSTP/55}
\end{flushright}
\vskip 0.5 truecm
\begin{center}
{\Large{\bf Anomalies, Local Counter Terms and Bosonization}%
\footnote{A contribution to the Hidenaga Yamagishi commemorative
volume of Physics Reports, edited by E.~Witten and I.~Zahed.}}
\end{center}
\vskip .5 truecm
\centerline{\bf Kazuo Fujikawa }
\vskip .4 truecm
\centerline {\it Department of Physics, University of Tokyo}
\centerline {\it Bunkyo-ku, Tokyo 113, Japan}
\vskip 0.5 truecm
\centerline{\bf Hiroshi Suzuki}
\vskip .4 truecm
\centerline {\it Department of Mathematical Sciences, Ibaraki University}
\centerline {\it Mito, 310-8512, Japan}

\makeatletter
\@addtoreset{equation}{section}
\def\theequation{\thesection.\arabic{equation}}
\makeatother
\setcounter{footnote}{0}

\begin{abstract}
We re-examine the issue of local counter terms in the analysis of quantum
anomalies. We analyze two-dimensional theories and show that the notion of
local counter terms need to be carefully defined depending on the physics
contents such as whether one is analyzing gauge theory or bosonization. It is
shown that a part of the Jacobian, which is apparently spurious and eliminated
by a local counter term corresponding to the mass term of the gauge field in
gauge theory, cannot be removed by a {\it local\/} counter term and plays a
central role by giving the kinetic term of the bosonized field in the context
of path integral bosonization.
\end{abstract}

\section{Introduction}
The quantum anomalies play a central role in modern field
theory~\cite{adler,jackiw,bertlmann}. In the analysis of quantum anomalies, it
is a folklore that the anomaly (or a part of anomaly), which can be cancelled
by local counter terms, has no physical meaning. However, it is
known~\cite{seo} that the apparently spurious Jacobian in path integral
formulation~\cite{fujikawa1}, which could be cancelled by a local counter term
corresponding to the mass term for the background gauge field, gives rise to
the kinetic term for the bosonized field in
bosonization~\cite{coleman}--\cite{abdalla}.

There are two basic issues involved when one discusses the local counter terms
in anomaly calculations:\\
1.~One has to decide if a relevant symmetry one is analyzing is really broken
by the quantum anomaly or not.\\
2.~One has to decide if all or parts of the anomalies evaluated in some
specific calculational schemes should be retained even if they appear to be
removed by adding suitable local counter terms to the starting Lagrangian.

In this paper, we present a detailed analysis of local counter terms and show
that the apparently spurious Jacobian in the context of gauge theory cannot be
removed by a {\it local\/} counter term in the context of bosonization.

\section{Anomalies in a two-dimensional model}
We work in the path integral formulation of quantum anomalies. We first briefly
review the calculation of the chiral anomaly in a theory which contains $U(1)$
gauge fields
\begin{equation}
   \int{\cal D}\bar\psi{\cal D}\psi\exp\left(i\int d^2x\,{\cal L}\right)
   =\int{\cal D}\bar\psi{\cal D}\psi
   \exp\left\{i\int d^2x\left[\bar\psi i\gamma^\mu
   (\partial_\mu-iV_\mu-iA_\mu\gamma_5)\psi\right]\right\}.
\end{equation}
The operator in Euclidean theory
\begin{equation}
   \Dslash\equiv\gamma^\mu(\partial_\mu-iV_\mu-iA_\mu\gamma_5)
\end{equation}
is not hermitian, and to define a hermitian operator we rotate the gauge field
into a pure imaginary variable
\begin{equation}
   A_\mu(x)\to iA_\mu(x)
\end{equation}
and rotate it back to the original field $A_\mu\to -iA_\mu$ after the
calculation~\cite{andrianov}. The operator after the rotation
\begin{equation}
   \Dslash\equiv\gamma^\mu(\partial_\mu-iV_\mu+A_\mu\gamma_5)
\end{equation}
becomes hermitian $\Dslash^\dagger=\Dslash$ in Euclidean sense and one can
define
\begin{eqnarray}
   &&\Dslash\varphi_n(x)=\lambda_n\varphi_n(x),\qquad
   \int d^2x\,\varphi_n^\dagger(x)\varphi_m(x)=\delta_{n,m},
\nonumber\\
   &&\psi(x)=\sum_na_n\varphi_n(x),\qquad
   \bar\psi(x)=\sum_n\bar b_n\varphi^\dagger(x).
\end{eqnarray}
In the path integral thus defined
\begin{equation}
   \int{\cal D}\bar\psi{\cal D}\psi\exp\left(\int d^2x\,{\cal L}\right)
   =\lim_{N\to\infty}\prod_{n=1}^Nd\bar b_nda_n
   \exp\left(\sum_{n=1}^Ni\lambda_n\bar b_na_n\right)
\end{equation}
the vector-like transformation
\begin{equation}
   \psi(x)\to\psi'(x)=e^{i\alpha(x)}\psi(x),\qquad
   \bar\psi(x)\to\bar\psi'(x)=\bar\psi(x)e^{-i\alpha(x)}
\end{equation}
gives rise to a vanishing Jacobian
\begin{equation}
   \ln J(\alpha)=-i\lim_{N\to\infty}\sum_{n=1}^N\int d^2x\left[
   \varphi^\dagger_n\alpha(x)\varphi_n
   -\varphi^\dagger_n\alpha(x)\varphi_n\right]=0
\end{equation}
and thus no anomaly. On the other hand, for the chiral transformation
\begin{equation}
   \psi(x)\to\psi'(x)=e^{i\alpha(x)\gamma_5}\psi(x),\qquad
   \bar\psi(x)\to\bar\psi'(x)=\bar\psi(x)e^{i\alpha(x)\gamma_5}
\end{equation}
the Jacobian is given by the master formula of quantum anomaly
\begin{eqnarray}
   \ln J_5(\alpha)&=&-2i\lim_{N\to\infty}\sum_{n=1}^N
   \int d^2x\left[\varphi^\dagger_n\alpha(x)\gamma_5\varphi_n\right]
\nonumber\\
   &\equiv&-2i\lim_{M\to\infty}\int d^2x\sum_{n=1}^\infty
   \left[\varphi^\dagger_n\alpha(x)\gamma_5\exp(-\lambda^2_n/M^2)
   \varphi_n\right]
\nonumber\\
   &=&-2i\lim_{M\to\infty}\Tr\langle x|\alpha(x)\gamma_5
   \exp(-\Dslash^2/M^2)|x\rangle
\end{eqnarray}
where the trace is taken over the freedom of Dirac indices in addition to the
space-time integral.

When one defines $D_\mu=\partial_\mu-iV_\mu$ as a covariant derivative with
only the vector-like gauge field (by noting $\eta^{\mu\nu}=(-1,-1)$ in
Euclidean metric) we have
\begin{eqnarray}
   \Dslash^2&=&\gamma^\mu\gamma^\nu
   (\partial_\mu-iV_\mu-A_\mu\gamma_5)(\partial_\nu-iV_\nu+A_\nu\gamma_5)
\nonumber\\
   &=&\eta^{\mu\nu}(\partial_\mu-iV_\mu-A_\mu\gamma_5)
   (\partial_\nu-iV_\nu+A_\nu\gamma_5)
\nonumber\\
   &&\quad+{1\over2}[\gamma^\mu,\gamma^\nu](\partial_\mu-iV_\mu-A_\mu\gamma_5)
   (\partial_\nu-iV_\nu+A_\nu\gamma_5)
\nonumber\\
   &=&D^\mu D_\mu+(\partial^\mu A_\mu)\gamma_5
   +{1\over4}[\gamma^\mu,\gamma^\nu](-iF_{\mu\nu})
\nonumber\\
   &&\quad-A^\mu A_\mu+{1\over2}[\gamma^\mu,\gamma^\nu]
   [D_\mu A_\nu-A_\mu D_\nu]\gamma_5
\end{eqnarray}
where the differential operator in $(\partial^\mu A_\mu)$ acts only on $A_\mu$
and $F_{\mu\nu}=\partial_\mu V_\nu-\partial_\nu V_\mu$. If one recalls the
definitions of $\gamma_5$ and $\gamma^1=i\sigma^2$, $\gamma^2=i\sigma^1$
($\gamma^0=\sigma^1$) in terms of Pauli matrices in Euclidean theory and the
properties
\begin{eqnarray}
   &&\gamma_5\equiv(-i)\gamma^1\gamma^2=i\sigma^2\sigma^1=\sigma^3,
\nonumber\\
   &&\gamma_5{1\over4}[\gamma^\mu,\gamma^\nu]
   ={i\over2}\epsilon^{\mu\nu},\qquad\epsilon^{12}=1,
\end{eqnarray}
the last two terms in (2.11) do not contain the $\gamma$~matrices and do not
contribute to the anomaly in two-dimensions. Consequently, one can use
\begin{equation}
   \Dslash^2=D^\mu D_\mu+(\partial^\mu A_\mu)\gamma_5
   +{1\over4}[\gamma^\mu,\gamma^\nu](-iF_{\mu\nu})
\end{equation}
in the calculation of anomalies.

By this way one can evaluate the {\it integrable form\/} of
anomaly~\cite{wess}
\begin{eqnarray}
   \ln J_5(\alpha)
   &=&-2i\lim_{M\to\infty}\tr\int d^2x\,\alpha(x)\gamma_5
   \int{d^2k\over(2\pi)^2}\, e^{-ikx}\exp\left(-{\Dslash^2\over M^2}\right)
   e^{ikx}
\nonumber\\
   &=&-2i\lim_{M\to\infty}\tr\int d^2x\,\alpha(x)\gamma_5
   \int{d^2k\over(2\pi)^2}
\nonumber\\
   &&\quad\times e^{-ikx} \exp\left[-{D^\mu D_\mu+(\partial^\mu A_\mu)\gamma_5
   +{1\over4}[\gamma^\mu,\gamma^\nu](-iF_{\mu\nu})\over M^2}\right]e^{ikx}
\nonumber\\
   &&=2i\tr\int d^2x\,\alpha(x)\gamma_5\left[(\partial^\mu A_\mu)\gamma_5
   +{1\over4}[\gamma^\mu,\gamma^\nu](-iF_{\mu\nu})\right]
   \int{d^2k\over(2\pi)^2}\,e^{k^2}
\nonumber\\
   &&={2i\over4\pi}\tr\int d^2x\,\alpha(x)
   \left(\partial^\mu A_\mu+{1\over2}\epsilon^{\mu\nu}F_{\mu\nu}\right)
\end{eqnarray}
by applying the scale transformation $k_\mu\to Mk_\mu$ as in four-dimensional
theory and then retaining terms to the order $1/M^2$ in the expansion of the
exponential factor in powers of~$1/M$.

When one rotates back to the original variable $A_\mu\to-iA_\mu$ in (2.14) and
writes the result in Minkowski metric by noting
\begin{equation}
   \epsilon^{12}=1\to\epsilon^{10}=1
\end{equation}
one obtains the Jacobian after the trace over the Dirac indices
\begin{equation}
   \ln J_5(\alpha)={i\hbar\over\pi}\int d^2x\,\alpha(x)
   \left(\partial^\mu A_\mu+{1\over2}\epsilon^{\mu\nu}F_{\mu\nu}\right)
\end{equation}
where we added $\hbar$ to emphasize that this is the one-loop effect. In the
case of non-Abelian theory, this Jacobian is generalized to~\cite{bardeen}
\begin{eqnarray}
   \ln J_5(\alpha)&=&{i\hbar\over\pi}\tr\int d^2x\,\alpha^a(x)T^a
   \left[D^\mu A_\mu+{1\over2}\epsilon^{\mu\nu}\left(F_{\mu\nu}
   +i[A_\mu,A_\nu]\right)\right]
\nonumber\\
   &=&{i\hbar\over2\pi}\int d^2x\,\alpha^a(x) \left[(D^\mu A_\mu)^a
   +{1\over2}\epsilon^{\mu\nu}F_{\mu\nu}^a
   -{1\over2}\epsilon^{\mu\nu}f^{abc}A_\mu^bA_\nu^c\right].
\end{eqnarray}
If one follows the conventional wisdom, the first term in the above Jacobians,
for example $\partial^\mu A_\mu$ in (2.16), is eliminated by a local counter
term (corresponding to the mass term of the gauge field) and thus has no
physical meaning. However, this term does not diverge in two-dimensional theory
and in fact, it plays a central role by giving the kinetic term for the boson
field in bosonization to be discussed below.

\section{Anomalies and counter terms in gauge theory}
We examine the issue related to local counter terms in the context of a gauge
theory with a $\gamma_5$ coupling which is defined by a formal path integral
\begin{equation}
   \int[{\cal D}A_\mu]{\cal D}\bar\psi{\cal D}\psi
   \exp\left\{i\int d^2x\left[
   \bar\psi i\gamma^\mu(\partial_\mu-iA_\mu\gamma_5)\psi
   -{1\over4}(\partial_\mu A_\nu-\partial_\nu A_\mu)^2\right]\right\}
\end{equation}
where the path integral measure is not specified yet, and a suitable gauge
fixing term is included in the measure~$[{\cal D}A_\mu]$.

If one adopts the definition of the fermionic measure discussed in the
preceding section, which is related to the Pauli-Villars regularization, the
term $\partial^\mu A_\mu$ in the final expression of the Jacobian
$\ln J_5(\alpha)$ could be cancelled if one adds a suitable mass term
$A^\mu A_\mu$ for the gauge field $A_\mu$ in the starting theory
\begin{equation}
   {\cal L}=\bar\psi i\gamma^\mu
   (\partial_\mu-iA_\mu\gamma_5)\psi+{\hbar\over2\pi}A_\mu^2.
\end{equation}
In fact we have
\begin{eqnarray}
   &&Z(A_\mu+\partial_\mu\alpha)
\nonumber\\
   &&=\int{\cal D}\bar\psi{\cal D}\psi
   \exp\left(i\int d^2x\left\{\bar\psi i\gamma^\mu
   \left[\partial_\mu-i(A_\mu+\partial_\mu\alpha)\gamma_5\right]\psi
   +{\hbar\over2\pi}(A_\mu+\partial_\mu\alpha)^2\right\}\right)
\nonumber\\
   &&=\int{\cal D}\bar\psi'{\cal D}\psi'
   \exp\left(i\int d^2x\left\{\bar\psi'i\gamma^\mu
   \left[\partial_\mu-i(A_\mu+\partial_\mu\alpha)\gamma_5\right]\psi'
   +{\hbar\over2\pi}(A_\mu+\partial_\mu\alpha)^2\right\}\right)
\nonumber\\
   &&=\int{\cal D}\bar\psi{\cal D}\psi\,J_5(\alpha) \exp\left\{i\int
   d^2x\left[\bar\psi i\gamma^\mu (\partial_\mu-iA_\mu\gamma_5)\psi
   +{\hbar\over2\pi}(A_\mu+\partial_\mu\alpha)^2\right]\right\}
\nonumber\\
   &&=Z(A_\mu)
\end{eqnarray}
where the second equality states that the path integral is independent of the
naming of path integral variables, and we set
$\psi'(x)=\exp[i\alpha(x)\gamma_5]\psi(x)$. The Jacobian is given by
\begin{equation}
   J_5(\alpha)=\exp\left[{i\hbar\over\pi}\int d^2x\,
   \alpha(x)\partial^\mu A_\mu\right]
\end{equation}
and thus the gauge symmetry is preserved for an infinitesimal~$\alpha$.

In this sense we interpret that the gauge symmetry in the present
regularization
\begin{equation}
   \int[{\cal D}A_\mu]{\cal D}\bar\psi{\cal D}\psi
   \exp\left\{i\int d^2x\left[
   \bar\psi i\gamma^\mu(\partial_\mu-iA_\mu\gamma_5)\psi
   +{\hbar\over2\pi}A^2_\mu
   -{1\over4}(\partial_\mu A_\nu-\partial_\nu A_\mu)^2\right]\right\}
\end{equation}
is not broken by the anomaly.

The above Jacobian indicates the identity
\begin{eqnarray}
   &&\int{\cal D}\bar\psi{\cal D}\psi
   \exp\left\{i\int d^2x\left[\bar\psi i\gamma^\mu
   (\partial_\mu-iA_\mu\gamma_5)\psi
   +{\hbar\over2\pi}A_\mu^2\right]\right\}
\nonumber\\
   &&=\int{\cal D}\bar\psi'{\cal D}\psi'
   \exp\left\{i\int d^2x\left[\bar\psi'i\gamma^\mu
   (\partial_\mu-iA_\mu\gamma_5)\psi'
   +{\hbar\over2\pi}A_\mu^2\right]\right\}
\nonumber\\
   &&=\int{\cal D}\bar\psi{\cal D}\psi\,J_5(\alpha) \exp\left\{i\int
   d^2x\left[\bar\psi'i\gamma^\mu (\partial_\mu-iA_\mu\gamma_5)\psi'
   +{\hbar\over2\pi}A_\mu^2\right]\right\}
\end{eqnarray}
for the transformation
\begin{equation}
   \psi(x)\to\psi'(x)=e^{i\alpha(x)\gamma_5}\psi(x),\qquad
   \bar\psi(x)\to\bar\psi'(x)=\bar\psi(x)e^{i\alpha(x)\gamma_5}
\end{equation}
which gives the identity
\begin{equation}
   \partial_\mu[\bar\psi(x)\gamma^\mu\gamma_5\psi(x)]
   =-{\hbar\over\pi}\partial_\mu A^\mu(x).
\end{equation}
This shows that the operator non-conservation holds independent of the local
counter term. We can however define a conserved local operator
\begin{eqnarray}
   &&J_5^\mu(x)\equiv\bar\psi(x)\gamma^\mu\gamma_5\psi(x)
   +{\hbar\over\pi}A^\mu(x),
\nonumber\\
   &&\partial_\mu J_5^\mu(x)=0
\end{eqnarray}
which is a reflection of the fact that the local counter term can cancel the
anomaly.

In the above calculational scheme, one can maintain both of the fermion number
symmetry in an operator sense
\begin{equation}
   \partial_\mu[\bar\psi(x)\gamma^\mu\psi(x)]=0
\end{equation}
and chiral gauge symmetry in the sense of the generating functional $Z(A_\mu)$
which is made invariant by a local counter term or by the re-definition
of~$J_5^\mu$.

In two-dimensions, the connection between the vector and axial-vector currents
adds further constraints. One may {\it define\/} a new (gauge invariant)
fermion number current in the present regularization scheme by (by noting
$\gamma^\mu\gamma_5=-\epsilon^{\mu\nu}\gamma_\nu$ and
$\epsilon^{10}=1$ in strictly $d=2$ Minkowski metric)
\begin{eqnarray}
   J^\mu(x)&\equiv&-\epsilon^{\mu\nu}J_{5\,\nu}(x)
\nonumber\\
   &=&\bar\psi(x)\gamma^\mu\psi(x)-{\hbar\over\pi}\epsilon^{\mu\nu}A_\nu(x)
\end{eqnarray}
which satisfies an anomalous relation
\begin{equation}
   \partial_\mu J^\mu(x)=-{\hbar\over\pi}\epsilon^{\mu\nu}
   \partial_\mu A_\nu(x).
\end{equation}
The fermion number current, which is obtained from the anomaly-free gauge
current~$J^\mu_5(x)$, is not conserved.

This may be compared to the {\it covariant formulation\/} of the Jacobian (in
Euclidean metric)~\cite{fujikawa2}
\begin{equation}
   Z_{\rm cov}=\int[{\cal D}A_\mu]{\cal D}\bar\psi{\cal D}\psi
   \exp\left\{\int d^2x\left[\bar\psi i\gamma^\mu
   (\partial_\mu-iA_\mu\gamma_5)\psi
   -{1\over4}(\partial_\mu A_\nu-\partial_\nu A_\mu)^2\right]\right\}.
\end{equation}
The calculational scheme of covariant anomaly is specified by (in Euclidean
theory)
\begin{eqnarray}
   &&\Dslash=\gamma^\mu(\partial_\mu-iA_\mu\gamma_5)
   =\gamma^\mu(\partial_\mu-iA_\mu)\left({1+\gamma_5\over2}\right)
   +\gamma^\mu(\partial_\mu+iA_\mu)\left({1-\gamma_5\over2}\right),
\nonumber\\
   &&\Dslash^\dagger
   =\gamma^\mu(\partial_\mu-iA_\mu)\left({1-\gamma_5\over2}\right)
   +\gamma^\mu(\partial_\mu+iA_\mu)\left({1+\gamma_5\over2}\right),
\nonumber\\
   &&\Dslash^\dagger\Dslash
   =[\gamma^\mu(\partial_\mu-iA_\mu)]^2\left({1+\gamma_5\over2}\right)
   +[\gamma^\mu(\partial_\mu+iA_\mu)]^2\left({1-\gamma_5\over2}\right),
\nonumber\\
   &&\Dslash\Dslash^\dagger
   =[\gamma^\mu(\partial_\mu-iA_\mu)]^2\left({1-\gamma_5\over2}\right)
   +[\gamma^\mu(\partial_\mu+iA_\mu)]^2\left({1+\gamma_5\over2}\right),
\nonumber\\
   &&\Dslash^\dagger\Dslash\psi_n(x)=\lambda_n^2\psi_n(x),\qquad
   \Dslash\Dslash^\dagger\varphi_n(x)=\lambda_n^2\varphi_n(x),
\nonumber\\
   &&\psi(x)=\sum_na_n\psi_n(x),\qquad
   \bar\psi(x)=\sum_n\bar b_n\varphi_n^\dagger(x)
\end{eqnarray}
and the path integral and operators appearing there are specified by the
cut-off in terms of the gauge invariant $\lambda_n$ with $\lambda_n\geq 0$
\begin{eqnarray}
   Z_{\rm cov}&=&\int[{\cal D}A_\mu]\lim_{N\to\infty}\prod_{n=1}^Nd\bar b_nda_n
   \exp\left[i\sum_{n=1}^N\lambda_n\bar b_na_n
   -\int d^2x\,{1\over4}(\partial_\mu A_\nu-\partial_\nu A_\mu)^2 \right].
\end{eqnarray}

If one applies the calculational scheme of covariant anomaly, the fermion
number (in an operator sense) in this theory however contains an anomaly,
though the chiral gauge current in an operator sense is anomaly-free. The
Jacobian for the fermion number transformation
\begin{equation}
   \psi'(x)=e^{i\alpha(x)}\psi(x),\qquad
   \bar\psi'(x)=\bar\psi(x)e^{-i\alpha(x)}
\end{equation}
is defined by
\begin{eqnarray}
   \ln J(\alpha)&=&-i\Tr\left\{\alpha(x)\left[
   \exp(-\Dslash^\dagger\Dslash/M^2)-
   \exp(-\Dslash\Dslash^\dagger/M^2)\right]\right\}
\nonumber\\
   &=&-i\Tr\biggl[\alpha(x)\biggl(
   \exp\left\{-[\gamma^\mu(\partial_\mu-iA_\mu)]^2/M^2\right\}
   \left({1+\gamma_5\over2}\right)
\nonumber\\
   &&\qquad\qquad\quad
   +\exp\left\{-[\gamma^\mu(\partial_\mu+iA_\mu)]^2/M^2\right\}
   \left({1-\gamma_5\over2}\right)
\nonumber\\
   &&\qquad\qquad\quad
   +\exp\left\{-[\gamma^\mu(\partial_\mu-iA_\mu)]^2/M^2\right\}
   \left({1-\gamma_5\over2}\right)
\nonumber\\
   &&\qquad\qquad\quad
   +\exp\left\{-[\gamma^\mu(\partial_\mu+iA_\mu)]^2/M^2\right\}
   \left({1+\gamma_5\over2}\right)\biggr)\biggr]
\nonumber\\
   &=&-2i\Tr\left(\alpha(x)\gamma_5
   \exp\left\{-[\gamma^\mu(\partial_\mu-iA_\mu)]^2/M^2\right\}\right)
\nonumber\\
   &=&\int d^2x\,\alpha(x)\,{i\hbar\over2\pi}
   \epsilon^{\mu\nu}(\partial_\mu A_\nu-\partial_\nu A_\mu)
\end{eqnarray}
and the Jacobian for the chiral transformation by
\begin{equation}
   \ln J_5(\alpha) =-i\Tr\left\{\alpha(x)\gamma_5
   \left[\exp(-\Dslash^\dagger\Dslash/M^2)
   +\exp(-\Dslash\Dslash^\dagger/M^2)\right]\right\}=0.
\end{equation}
We thus have the operator relations
\begin{eqnarray}
   &&\partial_\mu[\bar\psi(x)\gamma^\mu\gamma_5\psi(x)]_{\rm cov}=0,
\nonumber\\
   &&\partial_\mu[\bar\psi(x)\gamma^\mu\psi(x)]_{\rm cov}
   =-{\hbar\over\pi}\epsilon^{\mu\nu}\partial_\mu A_\nu
\end{eqnarray}
which reproduce the results of the gauge invariant currents $J^\mu_5(x)$
and~$J^\mu(x)$ in (3.9) and~(3.12), respectively, in the regularization
corresponding to the Pauli-Villars regularization.

The physics contents of anomaly-free gauge theory with a $\gamma_5$ coupling
are thus compactly specified by the covariant regularization without a need to
introduce a local counter term; this is also the case in the four-dimensional
Weinberg-Salam theory.

\section{Anomalies and the bosonization of free fermions}
We next examine the issue related to local counter terms in the context of the
path integral bosonization. In the analysis of the bosonization of the fermion,
the fermion number by definition should have a well-defined anomaly-free
meaning. The fermion number thus replaces the gauge symmetry as a primary
symmetry in the preceding section. If one imposes this condition, the available
regularization is restricted essentially to the formulation corresponding to
the Pauli-Villars regularization but {\it without\/} a local counter term.

\subsection{Abelian bosonization}
We review the bosonization related to Abelian gauge
transformations~\cite{coleman}--\cite{furuya}. We thus start with
\begin{equation}
   \int{\cal D}\bar\psi{\cal D}\psi\exp\left(i\int d^2x\,{\cal L}\right)
   =\int{\cal D}\bar\psi{\cal D}\psi\exp\left
   \{i\int d^2x\left[\bar\psi i\gamma^\mu(\partial_\mu-iV_\mu-iA_\mu\gamma_5)
   \psi\right]\right\}
\end{equation}
where the path integral is specified by the regularization in Section~2. There
are two alternative ways to proceed to the bosonization.

We first start with
\begin{equation}
   e^{iW(v_\mu)}=\int{\cal D}\bar\psi{\cal D}\psi
   \exp\left[i\int d^2x\left(\bar\psi i\gamma^\mu\partial_\mu\psi
   +v_\mu\bar\psi\gamma^\mu\psi\right)\right].
\end{equation}
In this setting any {\it local\/} counter term should be expressed as a local
polynomial in~$v_\mu$.

Now we observe that the vector field in two-dimensional space-time contains two
independent components and it can be decomposed into two arbitrary real
functions~$\alpha$ and~$\beta$ as
\begin{equation}
   v_\mu(x)=\partial_\mu\alpha(x)+\epsilon_{\mu\nu}\partial^\nu\beta(x).
\end{equation}
After this decomposition, we can write (4.2) as
\begin{equation}
   e^{iW(v_\mu)}=\int{\cal D}\bar\psi{\cal D}\psi \exp\left\{i\int d^2x
   \left[\bar\psi i\gamma^\mu
   (\partial_\mu-i\partial_\mu\alpha-i\partial_\mu\beta\gamma_5)\psi
   \right]\right\}
\end{equation}
by noting $\epsilon_{\mu\nu}\gamma^\mu=\gamma_\nu\gamma_5$, which is valid in
strictly $d=2$ Minkowski metric. We extract the functions $\alpha$ and~$\beta$
as integrated Jacobians associated with the transformations of integration
variables $\psi$ and~$\bar\psi$. Using the results of Section~2 for the
infinitesimal transformations
\begin{eqnarray}
   &&\psi(x)\to\psi'(x)=\exp\left\{i\left[\delta\alpha(x)
   +\delta\beta(x)\gamma_5\right]\right\}\psi(x),
\nonumber\\
   &&\bar\psi(x)\to\bar\psi'(x)=\bar\psi(x)
   \exp\left\{i\left[-\delta\alpha(x)+\delta\beta(x)\gamma_5\right]\right\},
\end{eqnarray}
we have the Jacobians in (4.4)
\begin{eqnarray}
   &&\ln J(\delta\alpha)=0,
\nonumber\\
   &&\ln J(\delta\beta)={i\over\pi}\int d^2x\,\delta\beta(x)
   (\partial^\mu A_\mu+\epsilon^{\mu\nu}\partial_\mu V_\nu),
\end{eqnarray}
where $V_\mu=\partial_\mu\alpha$ and $A_\mu=\partial_\mu\beta$. Using these
Jacobians, the $\alpha$ and $\beta$ dependences in the action are extracted as
\begin{equation}
   e^{iW(v_\mu)}=\exp[i\Gamma(v_\mu)]\int{\cal D}\bar\psi{\cal D}\psi
   \exp\left[i\int d^2x\,(\bar\psi i\gamma^\mu\partial_\mu\psi)\right]
\end{equation}
where $\Gamma(v_\mu)$ stands for the integrated Jacobian (or anomaly), and it
has the form in Minkowski metric
\begin{eqnarray}
   i\Gamma(v_\mu)&=&{i\over\pi}\int d^2x\int_0^1ds\,
   \beta\partial^\mu(1-s)\partial_\mu\beta
\nonumber\\
   &=&{i\over\pi}\int d^2x
   \left(-{1\over2}\partial^\mu\beta\partial_\mu\beta\right).
\end{eqnarray}
The derivation of this integrated Jacobian proceeds as follows: One first
eliminates the component $\alpha$ by a vector-like transformation, which is
anomaly free and thus without any Jacobian. One then performs an infinitesimal
transformation parametrized by $ds\beta$ in the intermediate result where the
axial-vector component $\beta$ is partially extracted and given by
$A_\mu=(1-s)\partial_\mu\beta$ and $V_\mu=0$ in the general formula (4.6) of
the Jacobian. One then obtains the above formula $\Gamma(v_\mu)$ by integrating
over the parameter $s$ of infinitesimal transformations.

We can also write
\begin{equation}
   e^{iW(v_\mu)}=\int{\cal D}\xi\exp\left[
   {i\over\pi}\int d^2x\left({1\over2}\partial^\mu\xi\partial_\mu\xi\right)
   \right]\exp[i\Gamma(v_\mu)]
\end{equation}
since the absolute normalization of path integral does not matter in the
definition of~$W(v_\mu)$. We next shift the variable $\xi\to\xi+\beta$ and note
the ``translational invariance'' of the path integral measure
${\cal D}(\xi+\beta)={\cal D}\xi$. We then have
\begin{eqnarray}
   e^{iW(v_\mu)}&=&\int{\cal D}\xi\exp\left[
   {i\over\pi}\int d^2x\left({1\over2}\partial^\mu\xi\partial_\mu\xi
   -{1\over2}\partial^\mu\beta\partial_\mu\beta\right)\right]
\nonumber\\
   &=&\int{\cal D}\xi\exp\left[
   {i\over\pi}\int d^2x\left({1\over2}\partial^\mu\xi\partial_\mu\xi
   -v_\mu\epsilon^{\mu\nu}\partial_\nu\xi\right)\right].
\end{eqnarray}
In deriving the last line, we used
$\partial^\mu\partial_\mu\beta=\epsilon^{\mu\nu}\partial_\mu v_\nu$.

This relation shows that the theory of a free Dirac fermion $\psi$ and the
theory of a free real Bose field $\xi$ define the identical generating
functional $W(v_\mu)$ of Green's functions. By differentiating the generating
functional with respect to the source field $v_\mu$ twice and then setting
$v_\mu=0$, we obtain
\begin{equation}
   \langle T^*\bar\psi(x)\gamma^\mu\psi(x)\bar\psi(y)\gamma^\nu\psi(y)\rangle
   =\left({1\over\pi}\right)^2
   \langle T^*\epsilon^{\mu\alpha}\partial_\alpha\xi(x)
   \epsilon^{\nu\beta}\partial_\beta\xi(y)\rangle
\end{equation}
which shows the equivalence of the correlation functions described by a Dirac
fermion and a real boson in $d=2$ dimensional space-time.

An alternative way to the bosonization starts with (4.1) and we consider (with
$A_\mu=\partial_\mu\xi$)
\begin{equation}
   Z(\xi,V_\mu)=\int{\cal D}\bar\psi{\cal D}\psi\exp\left\{
   i\int d^2x\left[\bar\psi i\gamma^\mu
   (\partial_\mu-i\partial_\mu\xi\gamma_5-iV_\mu)\psi\right]\right\}
\end{equation}
and the chiral transformation
\begin{equation}
   \psi(x)=\exp[i\xi(x)\gamma_5]\psi'(x),\qquad
   \bar\psi(x)=\bar\psi'(x)\exp[i\xi(x)\gamma_5].
\end{equation}
The $\xi$ dependence of the action is then removed
\begin{eqnarray}
   Z(\xi,V_\mu)&=&\exp[i\Gamma(V_\mu,\xi)]
   \int{\cal D}\bar\psi'{\cal D}\psi'\exp\left\{
   i\int d^2x\left[\bar\psi'i\gamma^\mu
   (\partial_\mu-iV_\mu)\psi'\right]\right\}
\nonumber\\
   &=&\exp[i\Gamma(V_\mu,\xi)]\int{\cal D}\bar\psi{\cal D}\psi
   \exp\left\{i\int d^2x\left[\bar\psi i\gamma^\mu
   (\partial_\mu-iV_\mu)\psi\right]\right\}
\end{eqnarray}
where $\Gamma(V_\mu,\xi)$ stands for the integrated Jacobian (or anomaly), and
it has the form in Minkowski metric
\begin{eqnarray}
   i\Gamma(V_\mu,\xi)&=&{i\over\pi}\int d^2x\int_0^1ds\,\xi
   \left[\partial^\mu(1-s)\partial_\mu\xi
   +{1\over2}\epsilon^{\mu\nu}F_{\mu\nu}\right]
\nonumber\\
   &=&{i\over\pi}\int d^2x \left({1\over2}\xi\partial^\mu\partial_\mu\xi
   +{1\over2}\xi\epsilon^{\mu\nu}F_{\mu\nu}\right)
\nonumber\\
   &=&{i\over\pi}\int d^2x \left(-{1\over2}\partial^\mu\xi\partial_\mu\xi
   +{1\over2}\xi\epsilon^{\mu\nu}F_{\mu\nu}\right).
\end{eqnarray}
The derivation of this formula proceeds as follows: One performs an
infinitesimal transformation parametrized by $ds\xi$ in the intermediate result
where $\xi$ (and axial-vector gauge field) is given by
$A_\mu=(1-s)\partial_\mu\xi$. One then uses (2.16) and obtains the above
formula by integrating over the parameter $s$ of infinitesimal gauge
transformations.

We next note that the gauge field in two-dimensional space-time contains two
independent components and it can be decomposed into two arbitrary real
functions $\alpha$ and $\beta$ as
\begin{equation}
   V_\mu(x)=\partial_\mu\alpha(x)+\epsilon_{\mu\nu}\partial^\nu\beta(x).
\end{equation}
After this decomposition, we can write
\begin{equation}
   \gamma^\mu V_\mu(x)=\gamma^\mu\partial_\mu\alpha(x)
   +\gamma^\mu\gamma_5\partial_\mu\beta(x)
\end{equation}
by using the relations $\gamma^0=\sigma^1$, $\gamma^1=i\sigma^2$,
$\gamma_5=\sigma^3$ valid in Minkowski space-time. Consequently, we can write
(4.12) as
\begin{eqnarray}
   Z(\xi,V_\mu)&=&\int{\cal D}\bar\psi{\cal D}\psi
   \exp\left\{i\int d^2x\left[\bar\psi i\gamma^\mu
   (\partial_\mu-i\partial_\mu\xi\gamma_5-i\partial_\mu\alpha
   -i\gamma_5\partial_\mu\beta)\psi\right]\right\}
\nonumber\\
   &=&\int{\cal D}\bar\psi{\cal D}\psi\exp\left(
   i\int d^2x\left\{\bar\psi i\gamma^\mu
   [\partial_\mu-i\partial_\mu(\xi+\beta)\gamma_5]\psi\right\}\right)
\end{eqnarray}
after the elimination of $\alpha$ by a change of variables
$\psi(x)\to\exp[i\alpha(x)]\psi(x)$ and
$\bar\psi(x)\to\bar\psi(x)\exp[-i\alpha(x)]$, which is free of anomaly.

We thus have
\begin{eqnarray}
   \int{\cal D}\xi\,Z(\xi,V_\mu)
   &=&\int{\cal D}\xi{\cal D}\bar\psi{\cal D}\psi
   \exp\left(i\int d^2x\left\{\bar\psi i\gamma^\mu
   [\partial_\mu-i\partial_\mu(\xi+\beta)\gamma_5]\psi\right\}\right)
\nonumber\\
   &=&\int{\cal D}\xi{\cal D}\bar\psi{\cal D}\psi\exp\left\{
   i\int d^2x\left\{\bar\psi i\gamma^\mu
   (\partial_\mu-i\partial_\mu\xi\gamma_5)\psi\right]\right\}
\nonumber\\
   &=&\int{\cal D}\xi\,Z(\xi,0)
\end{eqnarray}
by noting ${\cal D}\xi={\cal D}(\xi+\beta)$. When one combines this expression
with the formula (4.14) which contains $\Gamma(V_\mu,\xi)$, one concludes
\begin{eqnarray}
   &&\int{\cal D}\xi\,\exp[i\Gamma(V_\mu,\xi)]
   \int{\cal D}\bar\psi{\cal D}\psi
   \exp\left\{i\int d^2x\left[\bar\psi i\gamma^\mu(\partial_\mu-iV_\mu)\psi
   \right]\right\}
\nonumber\\
   &&=\hbox{independent of $V_\mu$}.
\end{eqnarray}

We now recall that the generating functional of connected Green's functions
$W(V_\mu)$, when the variable $V_\mu$ is regarded as the Schwinger's source
function, is defined by
\begin{equation}
   e^{iW(V_\mu)}=\int{\cal D}\bar\psi{\cal D}\psi
   \exp\left[ i\int d^2x\left(\bar\psi i\gamma^\mu\partial_\mu\psi
   +V_\mu\bar\psi\gamma^\mu\psi\right)\right].
\end{equation}
This expression combined with (4.20) gives rise to
\begin{equation}
   e^{-iW(V_\mu)}=\int{\cal D}\xi \exp\left[-{i\over\pi}\int d^2x
   \left({1\over2}\partial^\mu\xi\partial_\mu\xi
   -V_\mu\epsilon^{\mu\nu}\partial_\nu\xi\right)\right]
\end{equation}
by noting that the absolute normalization of the path integral is immaterial in
the definition of $W(V_\mu)$. On the right-hand side of this path integral, the
Feynman's $i\epsilon$ prescription appears with a reversed signature in accord
with the Wick rotation in this formula which is defined by $x^0\to+ix^2$
instead of the conventional $x^0\to-ix^2$. The hermitian conjugate of this
relation gives
\begin{equation}
   e^{iW(V_\mu)}=\int{\cal D}\xi\exp\left[{i\over\pi}\int d^2x
   \left({1\over2}\partial^\mu\xi\partial_\mu\xi
   -V_\mu\epsilon^{\mu\nu}\partial_\nu\xi\right)\right]
\end{equation}
which is justified in the expansion in terms of the source
function~$V_\mu$.\footnote{This procedure is justified only for tree diagrams.}

These relations (4.21) and (4.23) show that the theory of a free Dirac fermion
$\psi$ and the theory of a free real Bose field $\xi$ define the identical
generating functional $W(V_\mu)$ of Green's functions. We thus arrive at the
same conclusion as in the first method.

\subsection{Non-Abelian bosonization}

We next repeat the procedure (4.10) for the non-Abelian
bosonization~\cite{witten}--\cite{abdalla}. We define the generating functional
of Green's functions by
\begin{equation}
   e^{iW(v_-)}=\int{\cal D}\bar\psi{\cal D}\psi
   \exp\left[i\int d^2x\left(\bar\psi i\gamma^\mu\partial_\mu\psi
   +\bar\psi v_-\gamma^-\psi\right)\right],
\end{equation}
where the fermions belong to a representation of the group $SU(n)$ and the
source field~$v_-$ is Lie algebra valued, $v_-=v_-^aT^a$. We have used the
light-cone coordinates defined by
\begin{equation}
   A_\pm=A_0\pm A_1,\qquad A^\pm={1\over2}(A^0\pm A^1).
\end{equation}
We note the relation valid in strictly $d=2$
\begin{equation}
   \gamma^-={1\over2}(\gamma^0-\gamma^1)=\gamma^0{1+\gamma_5\over2}
   =-\gamma^1{1+\gamma_5\over2}
\end{equation}
and thus
\begin{equation}
   v_-\gamma^-=(v_0-v_1){1\over2}(\gamma^0-\gamma^1)
   =\gamma^\mu v_\mu{1+\gamma_5\over2}.
\end{equation}
In two-dimensional space-time, the two independent components of~$v_\mu$ can be
written as
\begin{eqnarray}
   &&-iv_-(x)=\exp[ih(x)]\partial_-\exp[-ih(x)]
   \equiv U(h)^\dagger\partial_-U(h),
\nonumber\\
   &&-iv_+(x)=\exp[il(x)]\partial_+\exp[-il(x)]
   \equiv U(l)^\dagger\partial_+U(l),
\end{eqnarray}
where $h=h^aT^a$ and~$l=l^aT^a$.

By using the above relations, the generating functional is written as
\begin{equation}
   e^{iW(v_-)}=\int{\cal D}\bar\psi{\cal D}\psi
   \exp\left(i\int d^2x\left\{\bar\psi i\gamma^\mu\left[\partial_\mu
   +U(h)^\dagger\partial_\mu U(h){1+\gamma_5\over2}\right]\psi\right\}\right).
\end{equation}
Formally, this can be regarded as a non-Abelian chiral gauge theory in which
vector and axial gauge fields are given by
$V_\mu=A_\mu=iU(h)^\dagger\partial_\mu U(h)/2$.

In (4.29), we repeatedly apply an infinitesimal transformation of integration
variables,
\begin{equation}
   \psi(x)\to\psi'(x)=\exp[idsh(x)(1+\gamma_5)/2]\psi,\quad
   \bar\psi(x)\to\bar\psi'(x)=\bar\psi\exp[idsh(x)(-1+\gamma_5)/2]
\end{equation}
and integrate the resulting Jacobian by using (2.17) with respect to the
parameter~$s$. In the intermediate steps, the gauge fields are given by
$V_\mu(s)=A_\mu(s)=iU(h,s)^\dagger\partial_\mu U(h,s)/2$ where
$U(h,s)=e^{-i(1-s)h}$. This operation then extracts the $v_-$ dependence from
the action of the path integral. The result is
\begin{equation}
   e^{iW(v_-)}=\exp[i\Gamma(v_-)]\int{\cal D}\bar\psi{\cal D}\psi
   \exp\left[ i\int d^2x\left(\bar\psi i\gamma^\mu\partial_\mu\psi\right)
   \right]
\end{equation}
where
\begin{equation}
   i\Gamma(v_-)={i\over2\pi}\tr\int d^2x\int_0^1ds\,h\left(D^\mu A_\mu(s)
   +{1\over2}\epsilon^{\mu\nu}
   \left\{F_{\mu\nu}(s)+i[A_\mu(s),A_\nu(s)]\right\}\right).
\end{equation}
Various factors in this expression are given explicitly as
\begin{eqnarray}
   &&h=-iU(h,s)^\dagger\partial_s U(h,s),
\nonumber\\
   &&D^\mu A_\mu(s)=\partial^\mu A_\mu(s)-i[V^\mu(s),A_\mu(s)]
   ={i\over2}\partial^\mu[U(h,s)^\dagger\partial_\mu U(h,s)],
\nonumber\\
   &&F_{\mu\nu}(s)+i[A_\mu(s),A_\nu(s)]
   =\partial_\mu V_\nu(s)-\partial_\nu V_\mu(s)-i[V_\mu(s),V_\nu(s)]
   +i[A_\mu(s),A_\nu(s)]
\nonumber\\
   &&={i\over2}\partial_\mu[U(h,s)^\dagger\partial_\nu U(h,s)]
   -{i\over2}\partial_\nu[U(h,s)^\dagger\partial_\mu U(h,s)].
\end{eqnarray}
Plugging these into (4.32) and using
$\partial^\mu(U^\dagger\partial_s U)=\partial_s(U^\dagger\partial^\mu U)
+[U^\dagger\partial_sU,U^\dagger\partial^\mu U]$, we find
\begin{equation}
   i\Gamma(v_-)=-i\Gamma_{\rm WZW}(U(h))
\end{equation}
where the Wess-Zumino-Witten action~$\Gamma_{\rm WZW}(U)$ is given by
\begin{equation}
   \Gamma_{\rm WZW}(U)
   ={1\over8\pi}\tr\int d^2x\,\partial^\mu U^\dagger\partial_\mu U
   -{1\over12\pi}\tr\int d^3x\,\epsilon^{\mu\nu\lambda}
   (U^\dagger\partial_\mu U)(U^\dagger\partial_\nu U)
   (U^\dagger\partial_\lambda U).
\end{equation}
In the second term on the right-hand side an additional third dimension has
been introduced by~$U(x^0,x^1,x^2)=U(h(x^0,x^1),s=1-x^2)$ (we have set
$\epsilon^{102}=1$). It can be verified that the WZW action defined above
satisfies the composition law
\begin{equation}
   \Gamma_{\rm WZW}(U(h)U)
   =\Gamma_{\rm WZW}(U(h))+\Gamma_{\rm WZW}(U)
   -{i\over2\pi}\tr\int d^2x\,v_-U\partial^-U^\dagger.
\end{equation}

We have shown in (4.31) that
\begin{equation}
   e^{iW(v_-)}=\exp\left[-i\Gamma_{WZW}(U(h))\right]
\end{equation}
since the absolute normalization of the path integral does not matter in the
definition of~$W(v_-)$. We can re-write this relation by multiplying the
constant
\begin{equation}
   \int{\cal D}U\exp\left[i\Gamma_{WZW}(U)\right]
\end{equation}
to the right-hand side. We then change the integration variable as
$U\to U(h)U$. By using the gauge invariance of the integration measure,
${\cal D}(U(h)U)={\cal D}U$, and the composition law in (4.36), we recover the
non-Abelian bosonization formula~\cite{witten}
\begin{equation}
   e^{iW(v_-)}=\int{\cal D}U\exp\left[i\Gamma_{WZW}(U)
   -{i\over2\pi}\tr\int d^2x\,(v_-iU\partial^-U^\dagger)\right].
\end{equation}

\section{Local counter terms and bosonization}
In either method of the Abelian bosonization in the preceding section, the
term~$\partial^\mu A_\mu$ in the Jacobian factor in (4.6) or in (2.16) plays a
central role to give the kinetic term of the bosonic field~$\xi$ in (4.10) or
(4.23). If one eliminates the term~$\partial^\mu A_\mu$ by a local counter term
as in (3.2) or (3.5), the path integral bosonization as presented here does not
work.

One may argue for the absence of the counter term by saying that it is a
necessity to preserve the fermion number symmetry precisely in the
bosonization, which in turn specifies the regularization of the Jacobian. Once
the fermionic path integral is specified by a specific regularization, its full
contents should be retained in the faithful bosonization; a local counter term
could modify the contents of the fermion theory in general. The actual
situation is, however, more involved and subtle.

We thus start with the second method of Abelian bosonization in (4.12). One may
add a local counter term $A_\mu^2/(2\pi)$ to the action in the path integral as
\begin{equation}
   Z(\xi,V_\mu)=\int{\cal D}\bar\psi{\cal D}\psi
   \exp\left\{i\int d^2x\left[\bar\psi i\gamma^\mu
   (\partial_\mu-iA_{\mu}\gamma_5-iV_\mu)\psi
   +{1\over2\pi}A_\mu^2\right]\right\}
\end{equation}
where
\begin{equation}
   A_\mu\equiv\partial_\mu\xi.
\end{equation}
When one extracts the $\xi$ dependence from the fermionic sector by a change of
path integral variables, one arrives at
\begin{equation}
   Z(\xi,V_\mu)=\int{\cal D}\bar\psi{\cal D}\psi
   \exp\left\{i\int d^2x\left[\bar\psi i\gamma^\mu(\partial_\mu-iV_\mu)\psi
   +{1\over\pi}\xi\epsilon^{\mu\nu}\partial_\mu V_\nu\right]\right\}
\end{equation}
since a part of the Jacobian for the chiral transformation is cancelled by the
counter term. See (4.15). In this formula, the variable $\xi$ has no dynamical
degrees of freedom and thus the Jacobian factor has no physical meaning; it
simply corresponds to an addition of a c-number term coupled to the source
$V_\mu$. This shows that the second method of bosonization fails if one adds
the local counter term $A_\mu^2/(2\pi)$, and the second method is essentially
reduced to the first method of bosonization
\begin{equation}
   Z(v_\mu)=\int{\cal D}\bar\psi{\cal D}\psi
   \exp\left\{i\int d^2x\left[\bar\psi i\gamma^\mu(\partial_\mu-iv_\mu)\psi
   \right]\right\}.
\end{equation}
In this sense, the first method of bosonization is more intrinsic, though the
second method provides a clear physical picture, namely, $\xi$ as a
Nambu-Goldstone field derivatively coupled to the axial current and $V_\mu$
standing for the source field.

We thus analyze the first method of bosonization in (4.2). By repeating the
above procedure, one might attempt to eliminate the part of the
anomaly~$\partial^\mu A_\mu$ in (4.6), which is spurious from a view point of
gauge theory, by adding a local counter term~$A_\mu^2/(2\pi)$ to the original
Lagrangian. This counter term is written as $(\partial_\mu\beta)^2/(2\pi)$ in
the present context, and this counter term, if added, cancels the kinetic term
of~$\beta$ in (4.8) just as in the case we discussed above and invalidates the
bosonization.

We now examine this counter term in more detail. The term $A_\mu^2/(2\pi)$,
which is local in terms of the axial vector field~$A_\mu$ appearing in (2.16),
is actually not local in the context of (4.2). The local counter term in the
context of (4.2) should be expressed as a local polynomial of the source
field~$v_\mu$, as we have emphasized there. The would-be counter term is
written in terms of~$v_\mu$ as
\begin{eqnarray}
   {1\over2\pi}A_\mu^2&=&{1\over2\pi}(\partial_\mu\beta)^2
\nonumber\\
   &=&-{1\over2\pi}\epsilon^{\mu\nu}\partial_\mu v_\nu
   {1\over\partial_\rho\partial^\rho}
   \epsilon^{\alpha\beta}\partial_\alpha v_\beta
   =-{1\over8\pi}\epsilon^{\mu\nu}F_{\mu\nu}
   {1\over\partial_\rho\partial^\rho}
   \epsilon^{\alpha\beta}F_{\alpha\beta}
\end{eqnarray}
which is {\it not\/} local. It is thus not allowed to add this term as a
counter term to the definition of the original partition function~(4.2). This
term, if added, modifies the physical contents of the original fermionic theory
even in the {\it conventional\/} understanding of local counter terms.

We can give an identical reasoning for the absence of the local counter term in
non-Abelian bosonization in (4.35). From a view point of gauge theory, the
first term of (4.32) may be regarded as a spurious anomaly and one might add a
counter term~$\tr A_\mu^2/(2\pi)=(A_\mu^a)^2/(4\pi)$. This term, if added,
removes the kinetic term of the non-linear sigma model in (4.35) (note
$A_\mu^2=\partial^\mu U^\dagger\partial_\mu U/4$) and invalidates the above
non-Abelian bosonization. This term, however, is {\it not\/} local when
expressed in terms of the source field~$v_-$. In fact, noting the relation
\begin{equation}
   \partial_+v_-=i\partial_-[U(h)^\dagger\partial_+U(h)]
   +[v_-,U(h)^\dagger\partial_+U(h)]
   \equiv iD_-^v[U(h)^\dagger\partial_+U(h)]
\end{equation}
we find
\begin{equation}
   A_\mu^2=A_-A_+
   =-{1\over4}U(h)^\dagger\partial_-U(h)U(h)^\dagger\partial_+U(h)
   ={1\over4}v_-{1\over D_-^v}\partial_+v_-
\end{equation}
which is non-local in terms of~$v_-$.

To understand the meaning of the local counter term in bosonization better, it
is instructive to examine what happens if one adds a {\it local\/} term
$-z^2v_\mu^2/(2\pi)$ to the Lagrangian in (4.2)
\begin{equation}
   e^{iW(v_\mu)}=\int{\cal D}\bar\psi{\cal D}\psi
   \exp\left[i\int d^2x\left(\bar\psi i\gamma^\mu\partial_\mu\psi
   +v_\mu\bar\psi\gamma^\mu\psi-{z^2\over2\pi}v_\mu^2\right)\right]
\end{equation}
where $z$ stands for a real non-negative number $z\geq 0$.

We follow the first method of the path integral bosonization (4.2) and observe
\begin{equation}
   -\int d^2x\,{z^2\over 2\pi}v_\mu^2
   =-\int d^2x\,{z^2\over 2\pi} \left(\partial^\mu\alpha\partial_\mu\alpha
   -\partial^\mu\beta\partial_\mu\beta\right)
\end{equation}
for the replacement (4.3). As a consequence, we obtain the following generating
functional after extracting the variables $\alpha$ and $\beta$ from the
fermionic path integral
\begin{eqnarray}
   &&e^{iW(v_\mu)}
\\
   &&=\int{\cal D}\xi{\cal D}\eta\exp\left[
   {i\over\pi}\int d^2x \left({1\over2}\partial^\mu\xi\partial_\mu\xi
   +{1\over2}\partial^\mu\eta\partial_\mu\eta
   -{z^2\over2}\partial^\mu\alpha\partial_\mu\alpha
   -{1-z^2\over2}\partial^\mu\beta\partial_\mu\beta\right)\right]
\nonumber\\
   &&=\int{\cal D}\xi{\cal D}\eta\exp\left[{i\over\pi}\int d^2x
   \left({1\over2}\partial^\mu\xi\partial_\mu\xi
   -v_\mu\sqrt{1-z^2}\epsilon^{\mu\nu}\partial_\nu\xi
   +{1\over2}\partial^\mu\eta\partial_\mu\eta
   +v_\mu z\partial^\mu\eta\right)\right]
\nonumber
\end{eqnarray}
where we introduced two real scalar fields, $\xi$ (pseudoscalar) and $\eta$
(scalar), and made the appropriate shifts of the path integral variables $\xi$
and $\eta$ in the second line. We also used the relation
$\partial^\mu\partial_\mu\alpha =\partial^\mu v_\mu$.

The formulas (5.8) and (5.10), after differentiating twice with respect to
$v_\mu$ and then setting $v_\mu=0$, give rise to
\begin{eqnarray}
   &&\langle T^*\bar\psi(x)\gamma^\mu\psi(x)\bar\psi(y)\gamma^\nu\psi(y)\rangle
   +{i z^2\over\pi}\eta^{\mu\nu}\delta^2(x-y)
\nonumber\\
   &&=(1-z^2)\left({1\over\pi}\right)^2
   \langle T^*\epsilon^{\mu\alpha}\partial_\alpha\xi(x)
   \epsilon^{\nu\beta}\partial_\beta\xi(y)\rangle
   +z^2\left({1\over\pi}\right)^2
   \langle T^*\partial^\mu\eta(x)\partial^\nu\eta(y)\rangle
\nonumber\\
   &&=\int{d^2p\over(2\pi)^2}e^{-ip(x-y)}\,{i\over\pi}
   \left[(1-z^2)\left({p^\mu p^\nu\over p^2}-\eta^{\mu\nu}\right)
   +z^2{p^\mu p^\nu\over p^2}\right]
\end{eqnarray}
and it gives the same current correlation function as our previous
formula~(4.11)
\begin{eqnarray}
   \langle T^*\bar\psi(x)\gamma^\mu\psi(x)\bar\psi(y)\gamma^\nu\psi(y)\rangle
   &=&\left({1\over\pi}\right)^2
   \langle T^*\epsilon^{\mu\alpha}\partial_\alpha\xi(x)
   \epsilon^{\nu\beta}\partial_\beta\xi(y)\rangle
\nonumber\\
   &=&\int{d^2p\over(2\pi)^2}e^{-p(x-y)}\,
   {i\over\pi}\left({p^\mu p^\nu\over p^2}-\eta^{\mu\nu}\right)
\end{eqnarray}
which is natural since our specification of the path integral measure precisely
fixes the fermionic currents.

The vector current defined as a source current for $v_\mu$ in (5.8) is given by
\begin{equation}
   j^\mu(x)=\bar\psi(x)\gamma^\mu\psi(x)-{z^2\over\pi}v^\mu(x)
\end{equation}
and it satisfies in the present regularization of the fermionic path integral
\begin{equation}
   \partial_\mu j^\mu(x)=\partial_\mu[\bar\psi(x)\gamma^\mu\psi(x)]
   -{z^2\over\pi}\partial_\mu v^\mu(x)=-{z^2\over\pi}\partial_\mu v^\mu(x).
\end{equation}
The axial current defined by
\begin{equation}
   j_5^\mu(x)\equiv-\epsilon^{\mu\nu}j_\nu(x)
   =\bar\psi(x)\gamma^\mu\gamma_5\psi(x)+{z^2\over\pi}\epsilon^{\mu\nu}v_\nu(x)
\end{equation}
then satisfies
\begin{equation}
   \partial_\mu j_5^\mu(x)
   =-{1-z^2\over\pi}\epsilon^{\mu\nu}\partial_\mu v_\nu(x).
\end{equation}
When one understands the local counter term as a result of ``regularization
ambiguity'' (see Appendix for a detailed analysis), one presumes the existence
of a certain regularization which gives rise to the fermionic currents
satisfying the same relations as in (5.14) and~(5.16)
\begin{eqnarray}
   &&\partial_\mu[\bar\psi(x)\gamma^\mu\psi(x)]_{\rm reg}
   =-{z^2\over\pi}\partial^\mu v_\mu(x),
\nonumber\\
   &&\partial_\mu[\bar\psi(x)\gamma^\mu\gamma_5\psi(x)]_{\rm reg}
   =-{1-z^2\over\pi}\epsilon^{\mu\nu}\partial_\mu v_\nu(x)
\end{eqnarray}
and in strictly $d=2$
\begin{equation}
   [\bar\psi(x)\gamma^\mu\gamma_5\psi(x)]_{\rm reg}
   =-\epsilon^{\mu\nu}[\bar\psi(x)\gamma_\nu\psi(x)]_{\rm reg}.
\end{equation}
In this understanding of the local counter term, the above correlation function
(5.11) implies\footnote{It is confirmed that the regularization scheme
parametrized by $L_1=L_2=L_3=L_4=1-2z^2$ in (A.5)--(A.8) in Appendix gives rise
to this formula.}
\begin{eqnarray}
   &&\langle T^*[\bar\psi(x)\gamma^\mu\psi(x)]_{\rm reg}
   [\bar\psi(y)\gamma^\nu\psi(y)]_{\rm reg}\rangle
\nonumber\\
   &&=(1-z^2)\left({1\over\pi}\right)^2
   \langle T^*\epsilon^{\mu\alpha}\partial_\alpha\xi(x)
   \epsilon^{\nu\beta}\partial_\beta\xi(y)\rangle
   +z^2\left({1\over\pi}\right)^2
   \langle T^*\partial^\mu\eta(x)\partial^\nu\eta(y)\rangle
\nonumber\\
   &&=\int{d^2p\over(2\pi)^2}e^{-ip(x-y)}\,{i\over\pi}
   \left[(1-z^2)\left({p^\mu p^\nu\over p^2}-\eta^{\mu\nu}\right)
   +z^2{p^\mu p^\nu\over p^2}\right].
\end{eqnarray}

These correlation functions (5.12) and (5.19), when written in term of the
canonical $T$-product by applying the Bjorken-Johnson-Law (BJL)
prescription~\cite{bjorken},\footnote{The essence of the BJL prescription is
that the correlation function written in terms of the $T$-product
$\langle T A(x)B(y)\rangle$ is well defined and smooth near $x^0\sim y^0$, and
thus the large $p_0$ limit of the Fourier transform vanishes
\begin{equation}
   \lim_{p_0\to\infty}\int d^2x\,e^{ip(x-y)}\langle TA(x)B(y)\rangle=0.
\end{equation}
}
both give rise to the same result
\begin{eqnarray}
   \langle T\bar\psi(x)\gamma^\mu\psi(x)\bar\psi(y)\gamma^\nu\psi(y)\rangle
   &=&\langle T[\bar\psi(x)\gamma^\mu\psi(x)]_{\rm reg}
   [\bar\psi(y)\gamma^\nu\psi(y)]_{\rm reg}\rangle
\nonumber\\
   &=&\int{d^2p\over(2\pi)^2}e^{-ip(x-y)}\,
   {i\over\pi}\left({p^\mu p^\nu\over p^2}
   -\delta_0^\mu\delta_0^\nu\right).
\end{eqnarray}
This formula in terms of the $T$-product gives the ordinary
Goto-Imamura-Schwinger term and the current algebra (the Kac-Moody algebra)
\begin{eqnarray}
   &&[j^0(t,x),j^0(t,y)]=0,
\nonumber\\
   &&[j^0(t,x),j^1(t,y)]=-{i\over\pi}\delta'(x-y),
\nonumber\\
   &&[j^1(t,x),j^1(t,y)]=0,
\end{eqnarray}
where $j^\mu$ stands for the vector current given by either definition of the
current. These relations are derived by
\begin{eqnarray}
   &&\lim_{p_0\to\infty}p_0\int d^2x\,e^{ip(x-y)}\langle Tj^0(x)j^\nu(y)\rangle
\nonumber\\
   &&=\lim_{p_0\to\infty}\int d^2x\,e^{ip(x-y)}
   i\partial_0\langle Tj^0(x)j^\nu(y)\rangle
\nonumber\\
   &&=\lim_{p_0\to\infty}\int d^2x\,e^{ip(x-y)}
   i\left[\langle T\partial_0j^0(x)j^\nu(y)\rangle
   +\langle[j^0(x),j^\nu(y)]\rangle\delta(x^0-y^0)\right]
\nonumber\\
   &&=\int d^2x\,e^{ip_{1}(x-y)^{1}}
   i\langle[j^{0}(x),j^{\nu}(y)]\rangle\delta(x^{0}-y^{0})
\nonumber\\
   &&=\lim_{p_0\to\infty}p_0{i\over\pi}
   \left({p^0p^\nu\over p^2}-\delta_0^0\delta_0^\nu\right)
\nonumber\\
   &&={i\over\pi}p^1\delta_1^\nu
\end{eqnarray}
where we used the definition of the $T$-product, in particular,
\begin{equation}
   \lim_{p_0\to\infty}\int d^2x\,e^{ip(x-y)}
   \langle T\partial_0j^0(x)j^\nu(y)\rangle=0.
\end{equation}
The last relation in (5.22) is derived in a similar way. It is known that the
chiral anomaly and the Goto-Imamura-Schwinger term give the same effect in the
current algebra in $d=2$ dimensions.

{}From a view point of bosonization, the two limiting cases $z=0$ and~$z=1$
give a clear physical picture which introduces a single bosonic field. The
local counter term thus preserves the current algebra and the canonical
commutation relations of the bosonized fields, for example,
\begin{equation}
   [\Pi_\eta(t,x),\partial_y\eta(t,y)]=
   {1\over\pi}[\partial_t\eta(t,x),\partial_y\eta(t,y)]=i\delta'(x-y)
\end{equation}
for $z=1$. This is the content (in the context of bosonization) of the common
statement that the local counter term does not change physics.

The Jacobian, which could be eliminated by a local counter term in the context
of gauge theory, plays a fundamental role in the bosonization of fermion theory
by giving the kinetic term of the bosonized field. We have shown that those
apparently spurious Jacobians are not eliminated by {\it local\/} counter terms
in the context of bosonization, even if one follows the conventional definition
of local counter terms.

\section{Discussions}
We have analyzed the issue related to local counter terms in the evaluation of
the quantum anomaly. We have shown that this issue depends on the physical
situations: In the context of gauge theory the ordinary argument for the use of
local counter terms is justified. On the other hand, the use of local counter
terms in a naive sense spoils the path integral bosonization. We have shown
that the apparently local counter terms in the context of gauge theory are in
fact not local in the context of bosonization. This clearly shows that the
naive use of local counter terms changes the physics contents of the original
fermionic theory in bosonization.

Our analysis of the apparently local counter term in the context of
bosonization has a close analogue in the analysis of the Liouville action in
the quantization of string theory. One may start with the Lagrangian for the
bosonic string
\begin{equation}
   {\cal L}={1\over2}\sqrt{g}g^{\mu\nu}\partial_\mu X^a\partial_\nu X^a
\end{equation}
where the index $\mu$ of $x^\mu$ runs over $1$ and~$2$ and parameterizes the
world sheet, and the index $a=1\sim d$ where $d$ stands for the dimension of
the target space-time. In the conformal gauge condition of the world sheet
$g_{\mu\nu}(x)=\rho(x)\eta_{\mu\nu}=\exp[\sigma(x)]\eta_{\mu\nu}$, it is known
that the (carefully defined) path integral measure in
\begin{equation}
   \int d\mu\exp\left[-\int d^2x\left(
   {1\over2}\sqrt{g}g^{\mu\nu}\partial_\mu X^a \partial_\nu X^a\right)\right]
\end{equation}
gives rise to the Liouville action~\cite{polyakov2}--\cite{alvarez}
\begin{equation}
   d\mu\to d\mu\,\exp\left[-{26-d\over48\pi}\int d^2x\left(
   {1\over2}\partial^\mu\sigma\partial_\mu\sigma+{1\over2}m^2e^{\sigma}
   \right)\right]
\end{equation}
when one extracts the Weyl freedom $\rho$ dependence from the action. The first
term of the Liouville action (the kinetic term) appears to be eliminated by a
suitable local counter term. But it is known that the kinetic term is in fact
non-local when written in a gauge condition other than the conformal gauge and
has a structure
\begin{equation}
   -{26-d\over96\pi}\int d^2x\,\sqrt{g}R\frac{1}{\Box}R
\end{equation}
where $R$ stands for the Riemann scalar curvature in two-dimensional space; in
the conformal gauge one has $\sqrt{g}R=\partial^\mu\partial_\mu\ln\rho$. This
is analogous to our finding in (5.5). If this kinetic term should be freely
changed by a local counter term, one would not be able to distinguish a
critical string from a non-critical string by looking at the Weyl anomaly. On
the other hand, the second term in the Liouville action is written as the
cosmological term~$\sqrt{g}$ and its coefficient can be freely modified by a
suitable local counter term. Incidentally, the present treatment of the
cosmological term (for $d< 26$) suggests that the spurious Jacobian term, which
could be eliminated by a local counter term, should be retained with an
arbitrary but non-vanishing coefficient in a general context.\\

We dedicate the present paper to the memory of Hidenaga Yamagishi, whose
critical spirit contributed to the clarification of an anomaly-related issue of
the proton decay~\cite{yamagishi}.

\section*{Note added}
After submitting this paper for publication, related works~\cite{Banerjee:rv}
on the path integral bosonization in two dimensions came to our attention. The
authors of these papers analyze the problem from a view point of Bose symmetry
of vertices involved in Feynman diagrams. Their criterion is more restrictive
than ours in the sense that the addition of the local counter term in~(5.8),
for example, is not allowed in their scheme. The definition of the Bose
symmetry is somewhat subtle and (A.18), which may be regarded to be Bose
symmetric, does not work in the bosonization. In the path integral bosonization
the property $\gamma^\mu\gamma_5=-\epsilon^{\mu\nu}\gamma_\nu$ is used in an
essential way inside the action, and thus this symmetry need to be preserved in
the current level also. (A.18) does not respect this symmetry as is shown
in (A.19).

\appendix
\section{Classification of regularization schemes}
In this appendix, we present a classification of the various regularization
schemes of
\begin{equation}
   \int{\cal D}\bar\psi{\cal D}\psi
   \exp\left\{i\int d^2x
   \left[\bar\psi i\gamma^\mu(\partial_\mu-iV_\mu-iA_\mu\gamma_5)
   \psi\right]\right\}
\end{equation}
by assuming that the regularization is Lorentz covariant and characterized as
a regularization of current operators
\begin{equation}
   j_5^\mu(x)=\bar\psi(x)\gamma^\mu\gamma_5\psi(x),\quad
   j^\mu(x)=\bar\psi(x)\gamma^\mu\psi(x).
\end{equation}
We note that, in strictly $d=2$ dimensions, there is a relation between the
above two fermionic currents
\begin{equation}
   j_5^\mu(x)=-\epsilon^{\mu\nu}j_\nu(x)
\end{equation}
as a result of $\gamma^\mu\gamma_5=-\epsilon^{\mu\nu}\gamma_\nu$ if the
currents are uniformly regularized in terms of the regularized correlation
functions $\langle\psi_\alpha(x)\bar\psi_\beta(y)\rangle_{\rm reg}$.

Now we consider the expectation values of currents $j^\mu(x)$ and~$j_5^\mu(x)$
in the presence of the background axial and vector gauge fields, $A_\mu$
and~$V_\mu$. In terms of Feynman diagrams, it is given by one-loop diagrams
which contain a current and interaction vertices specified by
$i\int d^2y\,{\cal L}_{\rm int}=i\int d^2y\,j_5^\nu(y)A_\nu(y)$
and~$i\int d^2y\,{\cal L}_{\rm int}=i\int d^2y\,j^\nu(y)V_\nu(y)$. One-loop
diagrams with more than two insertions of interaction vertices are ultraviolet
convergent and they do not contribute to anomalies (they are independent of
regularization schemes, as long as the ultraviolet properties are
concerned).\footnote{In fact, it can be shown that correlation functions
corresponding to those diagrams identically vanish. Functions given in
(A.5)--(A.8) are the only non-trivial ones in this Abelian theory.} It is thus
sufficient to consider one-loop diagrams which contain only one interaction
vertex. These are given by current-current correlators such as
$\langle T^*j_5^\mu(x)j_5^\nu(y)\rangle_0$, where the subscript~$0$ indicates
that the correlator is evaluated in the absence of gauge fields.

There are two different ways to define current operators: The first is to
define the currents by a variation of the fermion variables
\begin{eqnarray}
   &&\psi(x)\to\exp\left[i\alpha(x)+i\beta(x)\gamma_5\right]\psi(x),
\nonumber\\
   &&\bar\psi(x)\to\bar\psi(x)\exp\left[-i\alpha(x)+i\beta(x)\gamma_5\right]
\end{eqnarray}
inside the action, which defines the Noether currents. The second way is to
define the currents as the source currents for the background gauge fields
$V_\mu(x)$ and $A_\mu(x)$. These two definitions of currents give rise to the
identical results in ordinary cases. But in the analysis of anomalies, they
generally give rise to different results, namely, $L_2\neq L_3$ in (A.6) and
(A.7) below; one may recall that the effects of the anomaly can be collected to
any one of the vertices, either the Noether current or the interaction vertex,
in the present context. Our convention in this appendix is that the first
current at the point~$x$ stands for the Noether current and the second current
at the point~$y$ stands for the source current coupled to the background gauge
field.

By applying the standard (un-regularized) Feynman rules and expanding
correlators with respect to the external momentum, we find (${\rm F.T.}$
denotes the Fourier transformation with $\int d^2x\,e^{ip(x-y)}$)
\begin{eqnarray}
   &&{\rm F.T.}\,\langle T^*j_5^\mu(x)j_5^\nu(y)\rangle_0
   ={i\over2\pi}\left[\left({2p^\mu p^\nu\over p^2}-\eta^{\mu\nu}\right)
   F(p^2)+\eta^{\mu\nu}L_1(p^2)\right],
\\
   &&{\rm F.T.}\,\langle T^*j^\mu(x)j_5^\nu(y)\rangle_0
   =-{i\over2\pi}\left[\left({2\epsilon^{\mu\alpha}p_\alpha p^\nu\over p^2}
   -\epsilon^{\mu\nu}\right)F(p^2)+\epsilon^{\mu\nu}L_2(p^2)\right],
\\
   &&{\rm F.T.}\,\langle T^*j_5^\mu(x)j^\nu(y)\rangle_0
   =-{i\over2\pi}\left[\left({2\epsilon^{\nu\alpha}p_\alpha p^\mu\over p^2}
   -\epsilon^{\nu\mu}\right)F(p^2)+\epsilon^{\nu\mu}L_3(p^2)\right],
\\
   &&{\rm F.T.}\,\langle T^*j^\mu(x)j^\nu(y)\rangle_0 ={i\over2\pi}\left[
   \left({2p^\mu p^\nu\over p^2}-\eta^{\mu\nu}\right)F(p^2)
   -\eta^{\mu\nu}L_4(p^2)\right].
\end{eqnarray}
In these expressions, the factor $F$ is given by a convergent integral and we
have
\begin{equation}
   F(p^2)=1
\end{equation}
for any regularization scheme. On the other hand, the factors~$L_1$, $L_2$,
$L_3$ and~$L_4$ are given by logarithmically divergent integrals and their
values depend on the regularization scheme. The tensor structures of these
divergent terms are determined by assuming the Lorenz covariance. The
factors~$L_1$, $L_2$, $L_3$ and~$L_4$ parameterize possible regularization
schemes which include the choice of local counter terms. For example, an
addition of the counter term~$A_\mu^2$ amounts to a change of~$L_1$ while
keeping $L_2$, $L_3$ and~$L_4$ invariant, though other factors could be
changed by imposing (A.3).

One might think that $L_2=L_3$ always holds in view of the symmetric
appearance of (A.6) and~(A.7). But they are in general different as we
explained above. In fact the covariant formulation in Section~3 treats (A.6)
and~(A.7) in a different manner and leads to $L_2\neq L_3$. The property
$L_2\neq L_3$ can be regarded as a breaking of the Bose symmetry among the
vertices.

{}From the above expressions, we have
\begin{eqnarray}
   &&{\rm F.T.}\,\partial_\mu^x\langle T^*j_5^\mu(x)j_5^\nu(y)\rangle_0
   ={1\over2\pi}[L_1(p^2)+1]p^\nu,
\nonumber\\
   &&{\rm F.T.}\,\partial_\mu^x\langle T^*j^\mu(x)j_5^\nu(y)\rangle_0
   =-{1\over2\pi}[L_2(p^2)-1]p_\mu\epsilon^{\mu\nu},
\nonumber\\
   &&{\rm F.T.}\,\partial_\mu^x\langle T^*j_5^\mu(x)j^\nu(y)\rangle_0
   ={1\over2\pi}[L_3(p^2)+1]p_\mu\epsilon^{\mu\nu},
\nonumber\\
   &&{\rm F.T.}\,\partial_\mu^x\langle T^*j^\mu(x)j^\nu(y)\rangle_0
   =-{1\over2\pi}[L_4(p^2)-1]p^\nu
\end{eqnarray}
and
\begin{eqnarray}
   {\rm F.T.}\,\langle T^*[j_5^\mu(x)+\epsilon^{\mu\alpha}j_\alpha(x)]
   j_5^\nu(y)\rangle_0
   ={i\over2\pi}[L_1(p^2)-L_2(p^2)]\eta^{\mu\nu},
\nonumber\\
   {\rm F.T.}\,\langle T^*[j_5^\mu(x)+\epsilon^{\mu\alpha}j_\alpha(x)]
   j^\nu(y)\rangle_0
   ={i\over2\pi}[L_3(p^2)-L_4(p^2)]\epsilon^{\mu\nu}.
\end{eqnarray}

Now using the above four factors $L_1$, $L_2$, $L_3$ and~$L_4$, we can classify
various regularization schemes we encountered so far. The first category may be
called a ``consistent type'' and it is characterized by the
property~$L_2=L_3$. It thus preserves the Bose symmetry among two vertices. As
a possible choice, we may take $L_1=L_2=L_3=L_4=+1$. This choice corresponds to
the Pauli-Villars prescription in the path integral treatment. In fact, for
this choice, we have from (A.10)
\begin{equation}
   \partial_\mu\langle j^\mu(x)\rangle=0
\end{equation}
corresponding to (2.8), and
\begin{equation}
   \partial_\mu\langle j_5^\mu(x)\rangle
   =-{1\over\pi}[\partial_\mu A^\mu(x)+\epsilon^{\mu\nu}\partial_\mu V_\nu(x)]
\end{equation}
corresponding to (2.16). This choice thus breaks only the axial current
conservation. Also, from (A.11), we see that the relation~(A.3) holds in an
operator sense without ``anomaly''. In other words, if one requires the vector
current (which may be the fermion number current) conservation and the
relation~(A.3), one inevitably has $L_1=L_2=L_3=L_4=+1$.

Next, if one supplements the local counter term~$A_\mu^2$ to the above
prescription as in~(3.2), it eliminates the anomaly in the axial-axial channel.
This prescription is thus given by $L_1=-1\neq L_2=L_3=L_4=+1$. In fact, we
have $\partial_\mu\langle j^\mu(x)\rangle=0$ and
\begin{equation}
   \partial_\mu\langle j_5^\mu(x)\rangle
   =-{1\over\pi}\epsilon^{\mu\nu}\partial_\mu V_\nu(x).
\end{equation}
Here we are identifying the current $j_5^\mu$ with the source current of the
axial gauge field. This $j_5^\mu$ is thus equivalent to the current~$J_5^\mu$
in~(3.9). This prescription however somewhat peculiar because it leads to an
anomaly in the relation~(A.3):
\begin{equation}
   \langle j_5^\mu(x)+\epsilon^{\mu\nu}j_\nu(x)\rangle={1\over\pi}A^\mu(x).
\end{equation}
To remedy this anomaly, we may use the current $J^\mu$ in~(3.11) as the vector
current~$j^\mu$. This choice corresponds to $L_1=L_2=L_3=L_4=-1$ and we have
the relation~(A.3) and
\begin{equation}
   \partial_\mu\langle j_5^\mu(x)\rangle=0,\quad
   \partial_\mu\langle j^\mu(x)\rangle
   =-{1\over\pi}[\epsilon^{\mu\nu}\partial_\mu A_\nu(x)+\partial_\mu V^\mu(x)]
\end{equation}
which are ``dual'' to the relations in (A.12) and~(A.13). The axial current is
conserved and the anomalies appear only in the vector current.

The covariant formulation in Section~3 belongs to a different category of
regularization which may be termed as a ``covariant type''. This category is
characterized by the conservation of the current operators $j_5^\mu(y)$
and~$j^\mu(y)$ coupled to the gauge fields. Namely, one requires
\begin{eqnarray}
   &&{\rm F.T.}\,\partial_\nu^y\langle T^*j_5^\mu(x)j_5^\nu(y)\rangle_0
   =-{1\over2\pi}[L_1(p^2)+1]p^\mu=0,
\nonumber\\
   &&{\rm F.T.}\,\partial_\nu^y\langle T^*j^\mu(x)j_5^\nu(y)\rangle_0
   ={1\over2\pi}[L_2(p^2)+1]\epsilon^{\mu\nu}p_\nu=0,
\nonumber\\
   &&{\rm F.T.}\,\partial_\nu^y\langle T^*j_5^\mu(x)j^\nu(y)\rangle_0
   =-{1\over2\pi}[L_3(p^2)-1]\epsilon^{\mu\nu}p_\nu=0,
\nonumber\\
   &&{\rm F.T.}\,\partial_\nu^y\langle T^*j^\mu(x)j^\nu(y)\rangle_0
   ={1\over2\pi}[L_4(p^2)-1]p^\mu=0,
\end{eqnarray}
which implies~$L_1=L_2=-1\neq L_3=L_4=+1$. It is interesting that this
requirement completely fixes the regularization ambiguity. This covariant
regularization satisfies (A.3) without anomaly and
\begin{equation}
   \partial_\mu\langle j_5^\mu(x)\rangle
   =-{1\over\pi}\epsilon^{\mu\nu}\partial_\mu V_\nu(x),\quad
   \partial_\mu\langle j^\mu(x)\rangle
   =-{1\over\pi}\epsilon^{\mu\nu}\partial_\mu A_\nu(x)
\end{equation}
corresponding to (3.19) (in (3.19) $V_\mu$ is set to 0). These anomalies being
proportional to the $\epsilon$-tensor contain no spurious parts. This is an
advantage of the covariant type when it is applied to gauge theory. However,
when the system is not anomaly-free as gauge theory, this regularization breaks
the Bose symmetry (as is manifested by $L_2\neq L_3$) and cannot be used to
extract the effect of anomaly as a local functional, namely, as the Wess-Zumino
term. When there is no vector gauge field~$V_\mu$ as in~(3.13), the factors
$L_3$ and~$L_4$ are irrelevant and the covariant type gives an identical result
as $L_1=L_2=-1$ in the consistent type, as was noted in the text.

Incidentally, it is possible to ``transform'' the covariant type into the
consistent type~\cite{Banerjee:bu}. The prescription is that one first defines
a regularized effective action by integrating the covariantly regularized
currents multiplied by gauge fields with respect to the gauge fields along a
certain path such that $0\to A_\mu$ and $0\to V_\mu$. Then consistently
regularized currents are obtained by differentiating the effective action thus
defined with respect to gauge fields. It can be shown that this prescription
corresponds to $L_1=-1$, $L_2=L_3=0$ and~$L_4=+1$. Being $L_2=L_3$, the Bose
symmetry is restored and simultaneously anomalies have no spurious part:
\begin{equation}
   \partial_\mu\langle j_5^\mu(x)\rangle
   =-{1\over2\pi}\epsilon^{\mu\nu}\partial_\mu V_\nu(x),\quad
   \partial_\mu\langle j^\mu(x)\rangle
   =-{1\over2\pi}\epsilon^{\mu\nu}\partial_\mu A_\nu(x).
\end{equation}
As a cost of these properties, the relation~(A.3) has an anomaly with this
regularization
\begin{equation}
   \langle j_5^\mu(x)+\epsilon^{\mu\nu}j_\nu(x)\rangle
   ={1\over2\pi}[A^\mu(x)+\epsilon^{\mu\nu}V_\nu(x)].
\end{equation}

\end{document}